\documentclass[conference]{IEEEtran}

\usepackage{url}
\usepackage{cite}
\usepackage{amsmath,amssymb,amsfonts}
\usepackage{algorithmic}
\usepackage{graphicx}
\usepackage{textcomp}
\usepackage{epsfig}
\usepackage{epsf}
\usepackage{bmpsize}
\usepackage{boxedminipage}
\usepackage[table]{xcolor}
\usepackage{lscape}

\usepackage{ulem}

\newcommand\BlackCell[1]{%
  \multicolumn{1}{c|}{\cellcolor{black}\textcolor{white}{#1}}
}

\newcommand{\comment}[1]{}

\usepackage{placeins}
\usepackage{listings}
\usepackage{lipsum}

\begin{document}
%
\title{Is it a great Autonomous Trading Strategy or you are just fooling yourself}

\author{Murilo Sibrão Bernardini and Paulo André Lima de Castro \\
Autonomous Computational Systems Lab - LABSCA \\
Aeronautics Institute of Technology (ITA) \\
São José dos Campos, São Paulo, Brazil \\
murilosibrao@gmail.com, pauloac@ita.br}

\date{}
\maketitle

\begin{abstract}
In this paper, we propose a method for evaluating autonomous trading strategies that provides realistic expectations, regarding the strategy's long-term performance. This method addresses This method addresses many pitfalls that currently fool even experienced software developers and researchers, not to mention the customers that purchase these products. We present the results of applying our method to several famous autonomous trading strategies, which are used to manage a diverse selection of financial assets. The results show that many of these published strategies are far from being reliable vehicles for financial investment. Our method exposes the difficulties involved in building a reliable, long-term strategy and provides a means to compare potential strategies and select the most promising one by establishing minimal periods and requirements for the test executions. There are many developers that create software to buy and sell financial assets autonomously and some of them present great performance when simulating with historical price series (commonly called backtests). Nevertheless, when these strategies are used in real markets (or data not used in their training or evaluation), quite often they perform very poorly. The proposed method can be used to evaluate potential strategies. In this way, the method helps to tell if you really have a great trading strategy or you are just fooling yourself.

\end{abstract}



\section{Introduction}
\label{sec:intro}

Building autonomous agents for trading, sometimes called Algorithmic trading or trading robots has been the focus of much research and development in recent years. There are many published papers relating remarkable performance of the strategies in backtests, which are simulations that use historical data. We discuss some of these results briefly in Section~\ref{sec:work}. Nevertheless, when these strategies are used in actual markets, using current data that was not part of their training or evaluation, quite often they perform very poorly. The average results vary considerably but are usually disappointing. Some analysts have even argued that a positive backtesting performance may be taken to indicate a negative performance in real financial scenarios. 

 The goal of an investment manager, automated or human, is to find and acquire the most desirable set of assets that fits investor's preference profile. The manager acquires these assets through the submission of buy and sell orders to the stock market and such decisions are based on analyzes of a set of possible interesting assets. Obviously, with so many assets to choose from, each with a wealth of data to review, a machine learning approach, if not required. It is important to note that there are also ethical and legal implications that should be taken into consideration in the process of building automated trading strategies. We recommend Wellman and Rajan's paper~\cite{Wellman:17} about ethical issues in Autonomous Trading Agents.

The broad spectrum of AI techniques used in finance includes reinforcement Learning~\cite{Pereira:19,Stone:04}, multiagent systems~\cite{Castro:13,Wooldridge:18} complex networks~\cite{Zhao:18}, decision trees~\cite{Santos:18}, genetic algorithms~\cite{Stone:06}, random forests~\cite{Abdenur:19} to more recent approaches, like convolutional neural networks~\cite{Paiva:19} and  deep reinforcement learning~\cite{DRL_Portfolio:2019}. Recent studies have combined several techniques as LSTM NN and reinforcement learning~\cite{Lee:21}, while others have explored an ensemble of deep learning models in a multi-layer architecture~\cite{Carta:21}. There are many cases where the developers are successful in creating strategies with great performance results when backtesting with historical price series. In fact, some electronic platforms offer thousands of autonomous agents, or trading robots, which are also called bots or expert advisors. Their developers claim this software will be profitable in real financial markets~\cite{Metatrader:18}. These agents' strategies may be created with a simple idea such as moving averages or use complex machine learning schemes. To date, however, these strategies have usually performed poorly with  data not used in their training, despite all the hype.

The problem of creating a trading strategy with great performance in some previous known data set, but with bad performance in new data is well known in Machine Learning field and it is usually called \textbf{overfitting}. Nevertheless, this problem seems to be unknown or at least underestimated by many developers in autonomous trading. In truth, overfitting is more likely to occur in finance than in the traditional machine learning problems on which it is based, such as face recognition. Moreover, the consequences of a false positive can be far more disastrous in financial trading. False expectations are created that can mislead investors, putting their financial livelihood at stake. There is very little out there to protect investors from naive, incompetent, or even malicious software developers. Section~\ref{ssec:overf}, Overfitting in Finance, deals with this important issue.

In this paper, we propose a method for evaluating trading strategies that implements procedures that may applied regardless of the way the strategies were built or the technology they use. This method is called the (statistically) Significant Trading-Strategy Evaluation, or STSE.  It highlights many pitfalls that have been overlooked by software developers, including experienced machine learning experts, thus minimizing financial risks when using these strategies. We present the results of applying this method to several famous autonomous strategies used in foreign exchange assets and stock market. The results show that these strategies are far from being reliable vehicles of investment. Some periods of returns can be observed, but they are not predictable and probably happen by pure chance, rather than by the software having identified a real pattern in the input data. This fleeting return can be used maliciously by those selling the strategies, by their advertising great, but non-repetitive results, while obscuring negative results. The method we propose can be used to select among potential strategies, by establishing minimal periods and requirements as part of the test to avoid this seesaw in return results. Some autonomous approaches have been tried for dealing with derivative instruments~\cite{Sandholm:11}. We did not apply our method to any derivative instruments, which involve the future value of an underlying asset~\cite{Hull:06}, because of their price volatility~\cite{Hull:06}. Instead, we focused on foreign exchanges and stock assets. 

The remainder of this paper is organized as follows: In Section~\ref{sec:work}, we discuss some comparable proposals. In Section~\ref{sec:fin_env}, we present some distinct characteristics of  financial markets that make them an unique challenge for machine learning and artificial intelligence approaches. In Section~\ref{sec:STSE}, we propose our Significant Trading-Strategy Evaluation (STSE) method. We used it to evaluate several well known trading strategies currently on the market. The results are discussed in Section~\ref{sec:results}. Finally, we discuss some possible extensions and future study in Section~\ref{sec:conc}. We have also all necessary details to reproduce the experiments (Appendix~\ref{sec:setup}), which results are are shown in Section~\ref{sec:results}. 

We believe that reproducible results are important for research and that there is a significant reproducibility crisis in general. In fact, a recent survey~\cite{Baker:16} published in Nature journal showed that many researchers share that opinion. As pointed out by Stark~\cite{Stark:18}, researchers should provide enough information for others to repeat their experiments. We do that in appendix~\ref{sec:setup}. We also provide the source code that implements STSE freely at github~\cite{stse:21}.

\section{AI in Finance}
\label{sec:fin_env}

A financial agent is responsible for selecting, buying and selling financial assets in order to fulfill its investors requirements for risk and return, respecting the investor's horizon of investment. A financial autonomous agent is a software program that implements a trading strategy able to perform such activities more efficiently. 

The Financial Environment where these trading strategies are applied may be classified as: partially observable, sequential, stochastic, dynamic, continuous and multiagent, according to the Russell and Norvig taxonomy~\cite{Russell:13}. The environment is also strategic, in the sense that two active investors compete for a more accurate valuation of assets and their acts may change other economic agents' behavior. Russel and Norvig consider this the most complex environment class~\cite{Russell:13}. 

This taxonomy, however, does not really represents the whole complexity of the problem, because financial markets are not an independent and identically distributed (i.i.d) environment, as assumed in most of machine learning problems. Furthermore, financial environment have a low signal to noise ratio, which makes overfitting even more likely in machine learning approaches. Another aspect that should be addressed by any trading strategy, but is often forgotten, is that people do not have the same \textbf{level of acceptable risk}. Some investors may present a much stronger risk aversion than others. A trading agent must be aware of its investor's preferences, in order to trade appropriately. Investment errors caused by overfitting can also be very costly.

\subsection{Overfitting in Finance}
\label{ssec:overf}

Financial environments are not only stochastic environments but usually \textbf{non-stationary process} where the probability distribution do change along the time. Due to this non-stationary feature, a specific strategy may perform extremely well for a certain period of time and then perform very poorly thereafter. Moreover, \textbf{different assets may require different information and models}. For instance, oil companies may be very sensitive to changes of petrol prices, while probably it is not the same case for banks. In other terms, a specific asset price can be very correlated to some time series, but that time series may have no relevant correlation with another asset price.

Markets conditions may change abruptly leading to structural breaks in financial time series, which are  unexpected changes over time in the parameters of regression models. In fact, it can be checked by searching for structural breaks in financial time series. As stated by Andreou and Ghysels~\cite{Andreou:09}, there are abundant empirical evidence which shows that there are structural breaks in financial markets that may be caused by various economic events. When structural breaks occur the true model is changed; therefore, previous data does not describe the same process. It reduces the amount of useful data that can be used in fitting machine learning models. However, if that effect is not observed and outdated data is used, despite structural breaks, it may lead to models that do not generalize well in current market situations;or in other words, Overfitted models. Structural breaks do not usually occur in ML environments, which are traditionally considered to be independent and identically distributed (i.i.d). This hypothesis is so widely used in ML, that sometimes researchers forget that it is not valid for financial environments.  

Moreover, models with \textbf{high representational capacity}, such as deep learning approaches~\cite{Goodfellow:16}, have higher risks of overfitting, if they are not properly used. The problem is that such models have many parameters and ,therefore, they can fit many different solutions to the available data points. There is little chance that the learning algorithm will learn a  solution that generalizes well, i.e. it is not \textbf{overfitted},  when so many wildly different solutions exist~\cite{Goodfellow:16}.

In financial environments, it is hard to separate signal (real information) from the noise, because they are almost at the same level and, if the signal is clear, it will not be for long. Arbitrage, exploiting price differences of similar financial instruments on different markets, further decreases the signal-to-noise ratio, making it easier to confuse signal and noise. This confusion may also lead to \textbf{overfitted} models. 

Overfitting in finance leads to unrealistic expectation of the performance and, therefore, it can be used to mislead investors to bet on overfitted autonomous trading systems, which may be built by naive or even malicious developers~\cite{Prado:14}.

In fact, it is not rare to see promises of high return by trading strategy vendors in digital media. As pointed out by Prado~\cite{Prado:14}, without control for backtest overfitting, good backtest performance may be an indicator for negative future results.

Unfortunately, it is not easy to prove that a given model is not overfitted, despite the fact that there are many techniques in the ML field that help avoiding lack of generalization power~\cite{Witten:16,Goodfellow:16}. It is beyond the scope of this paper to address such general ML techniques to avoid overfitting and we are going to focus on specific features of financial environments. In fact, we believe it is fundamental to have a good understanding of the market faced by trading strategies in order to avoid problems. By that we mean the reasons that make overfitting more likely to happen in the financial environment and some common mistakes in building and evaluating trading strategies. We tackle these issues with the proposed method in Section~\ref{sec:STSE}.

\section{Related Work}
\label{sec:work}

There is a long history of studies related to evaluating portfolio management performance in the finance field~\cite{Brown:12}. However, there are some caveats when applying these methods to autonomous trading strategies. It is relatively simple to create new trading strategies using some machine learning technique and historical data or some model based approach with optimized parameters for a given market scenario. A developer needs to avoid some common pitfalls (1), select a meaningful measure (2) and be able to correctly pick among possible candidate strategies (3). We describe some related work that discuss these issues next. 

\subsection{Common pitfalls in evaluating trading strategies}
\label{sec:pitfalls}

{The complexity of the financial environment for autonomous agents, as discussed in Section~\ref{sec:fin_env}, hides common pitfalls for the evaluation of trading strategies.  We have noted that these problems have been overlooked in many studies as pointed out by other researchers~\cite{Luo:14}. The main pitfalls for evaluating trading strategies are the following:}

\begin{itemize}
	\item \textbf{Overfitting} allow bad trading strategies present good performance for certain periods of time, because they exploit false patterns that are unlikely to repeat in real markets and it can be easily happen in trading strategies.
	\item \textbf{Data leakage} can occur by using information that was not public at the moment the simulated decision was made. The time stamp for each data point must be observed and take into account release dates, distribution delays, and back-fill corrections~\cite{Luo:14}. One should also avoid training the model with testing data, or allow the leak of testing data to the training process. Some kind of leakage (or data-snooping) may also happen when a given set of data is used more than once for purposes of inference or model selection~\cite{Sullivan:99}. Data leakage may also happen using data points with very close time stamps in training and testing sets. See discussion  in~\cite{Prado:18} about overlapping and embargo (pgs. 105-109). 
	\item \textbf{Survivor bias} may occur by selecting to simulate only securities and companies that are still listed; therefore, ignoring all bankrupt and delisted securities in the process~\cite{Luo:14}.
	\item \textbf{Evaluating strategies without caring about risk}: In finance theory and finance institutions, a great deal of effort is devoted to to measure and control risk. Since the seminal work of Markowitz~\cite{Markowitz:52} that established modern portfolio theory, building diversified portfolio, in order to achieve high returns and mitigate risk is the norm in the finance field. It is surprising to observe that creators of financial autonomous strategies often underestimate the relevance of risk measurement and control and care only about return~\cite{Castro:13} (pg. 213). 
	\item \textbf{The Disregard transaction costs} produces strategies that look great in backtests but may bring negative returns systematically in real operation~\cite{Luo:14}. 
	\item \textbf{The lack of control for the impact of their own trading} is based on the assumption that the autonomous trading strategy will not have significant effect in the price or on the behaviour of others traders. This is really reasonable as long the capital controlled by the strategy is significantly smaller than the total volume traded. However, it is relevant to note in this context, that autonomous trading strategies are easy to copy usually. In fact, there are platforms that sell or rent autonomous strategies to third parties, which magnifies its impact. Furthermore, copying trading strategies or using other trades as primary information on other strategies are not new~\cite{etoro:10}. The idea of investors building and sharing investment strategies with other investors has been called social trading by some~\cite{Mcwaters:15} (pg. 127) and pointed out as a key trend in asset management~\cite{Mcwaters:15}. 
	\item \textbf{Oversimplifying shorting} may falsely enhance performance results. In a short position, the investor borrows stock and sells it with plans to buy it later. Taking a short position on spot markets requires finding a lender. The cost of lending and the amount available is generally unknown, and depends on several factors like relations, inventory and others~\cite{Luo:14}, but assuming it is zero is certainly a mistake.
\end{itemize}

\subsection{Sharpe ratio as Investment Skill measure}
\label{sec:sharpe}

Most of the work related to defining a meaningful performance evaluation for autonomous trading strategies rely on the Sharpe ratio. That includes those cited in this Section. In fact, the Sharpe ratio is well known and often used in evaluating fund performance. However, there are significant criticism about it. One criticism s is that it does not distinguish between upside and downside volatility. According to Rollinger and Hoffman~\cite{Rollinger:13}, high outlier returns could increase the value of the Sharpe ratio's denominator (standard deviation) more than the value of the numerator (return above the risk free return), thereby lowering the value of the ratio. Another criticism is that the assumption of normality in return distributions have been relaxed and the Sharpe ratio thus becomes a questionable tool for selecting optimal portfolios~\cite{Farinelli:08}. In fact, some indexes have been proposed as alternatives to the Sharpe ration, including the Sortino ratio~\cite{Sortino:96} and Modigliani index~\cite{Damodaran:10}. Despite the quality of these proposals, the Sharpe ratio "...definitely is the most widely used.." measure of risk-adjusted performance~\cite{Rollinger:13}(pg.42). Furthermore, as stated by Bailey and others~\cite{Bailey:12}(pg. 15) the Sharpe ratio, despite its known deficiencies, can provide evidence of investment skill since a proper track record length is observed.

\subsection{Picking among trading strategies}

Another notable related work defines a minimum track record length (or minimum backtest length) needed to avoid selecting the trading strategy with highest in-sample Sharpe Ratio among N independent strategies, but that has an expected out-of-sample Sharpe Ratio equal to zero~\cite{Prado:14}. The focus is avoiding selecting an overfitted strategy that performs well in the training (or in-sample) data, but performs poorly in real operations. Bailey and others~\cite{Prado:14} demonstrated the Theorem 1. It stresses that the backtest period must be longer as the number of independent strategies (N) grows in order to avoid the risk of selecting an overfitted strategy.

\textbf{Theorem 1}. The Minimum Backtest Period (MinBTL, in years) needed to avoid selecting a strategy with an in-sample Sharpe ratio of $E[max_{N}]$ among N independent strategies and with an expected out-of-sample Sharpe ratio of zero is given by Equation~\ref{eq:minbtl} (For demonstration, see~\cite{Prado:14}):
\begin{equation}
MinBTL < \frac{2*ln[N]}{E[max_{N}]^{2}}.
\label{eq:minbtl}
\end{equation}

In Section~\ref{sec:STSE}, we propose a method that incorporates such concerns when allows the evaluation of autonomous trading strategies, regardless of the way they were built. 

\section{The Significant Trading-Strategy Evaluation (STSE) Method - How to evaluate trading strategy performance}
\label{sec:STSE}

We propose a method called Significant Trading-Strategy Evaluation (STSE), that evaluates trading strategies in a statistically significant way regardless of the technologies used for building them. To be meaningful, an evaluation needs to cover a long enough time to avoid false positive results, which are unreproducible in the long term, as well as prevent data leakage from evaluation to training data and other  pitfalls (Section~\ref{sec:pitfalls}). There are four steps to our method.

\begin{enumerate}
\item Define Parameters:
\begin{enumerate}
	\item Financial parameters: set of assets, horizon of investment, and amount of capital
	\item Strategy parameters: according to picked approach and use available training data
\end{enumerate}
\item Verify conditions:
\begin{enumerate}
	\item Verify if the traded volume may lead to an underestimate of the strategy's impact on the market
	\item Verify if the strategy uses shorting, check if transaction costs have been tallied sufficiently
\end{enumerate}
\item \textit{Refine parameters} using historical prices (backtest) until a reasonable performance has been achieved 
\item \textit{Execute strategy} using real time simulation until achieving meaningful results (detailed in Section~\ref{sec:stse_fund})
\end{enumerate}

The first two phases, which define parameters and verify conditions, are quite direct, especially the first one. There are few details to keep in mind: \textbf{Underestimating impact of own trading.}. That pitfall is perhaps the least likely to happen, it is not even listed by~\cite{Luo:14}.  If your maximum traded volume is much smaller less than 10 base points than the average traded volume of the asset and the strategy is not used by others the impact of one's own trading is probably small enough to be unconsidered. To account for \textbf{Survivor bias.}, you need to state your criteria for selecting the target assets in a clear way and verify that if using that criteria and only information available in the beginning of the simulated time, you would select a problematic asset (delisting, bankruptcy, etc). To ensure that \textbf{Short operation costs} have been covered sufficiently you may use the maximum costs observed in the market for short operations similar tor the short operations executed by your strategy.

The third phase, refine parameters, uses the so-called backtest to fit parameters in order to improve performance. It should be carried out without any data leakage from test to training set by including an embargo period as discussed in Section~\ref{sec:pitfalls}. However, it is still possible to have some data leakage due to back-fill corrections or from avoiding assets that presented bad performance or negotiation interruptions (survivor bias). In fact, the best way to avoid data leakage of any kind and verify that the model represents patterns that are still valid in the market is by using real time simulation. That is the fourth step of STSE method.

Executing the strategy using real time simulation prevents data leakage, since each data point is out-of-sample and used for evaluation as soon as available. One could argue that a 'reasonable amount of time' is subjective. However, STSE uses a probabilistic approach that establishes the minimum time for passing a performance threshold with a given level of certainty. That amount of time is highly dependent on the strategy's observed performance. Great strategies may show skill briefly, while others require much more time and some others will never pass a given performance threshold even after many months or even years. In that case, the strategy is not a good option for the current market situation. Nevertheless since the financial environment is non-stationary, it is not possible to assure that such strategy will never present good performance in the future.

\subsection{STSE Fundamentals}
\label{sec:stse_fund}

STSE is based on the idea that you need to observe a trading strategy avoiding data leakage and mitigating overfitting risk for a long enough time. The amount of time is highly dependent on the trading strategy performance itself and the target performance threshold. This threshold is defined in terms of a minimum desired Sharpe Ratio for the trading strategy. Naturally, the observed Sharpe Ratio for any given trading strategy is highly dependent on market conditions, but it is possible to define confidence levels where the observed SR is equal or above the threshold even for non-Normal returns. In fact, Bailey and Prado~\cite{Bailey:12} showed that the probability that the observed Sharpe Ratio ($\widehat{SR}$) will be greater than a given Sharpe Ratio threshold ($SR^{*}$), can be given by Equation~\ref{eq:ProbSR}. 

In Equation~\ref{eq:ProbSR}, Z is the CDF (cumulative distribution function) of the Standard Normal distribution and $\widehat{\gamma_{3}}$ and $\widehat{\gamma_{4}}$ are the skewness and kurtosis, respectively.
\begin{equation}
\label{eq:ProbSR}
	Prob[\widehat{SR} > SR^{*}] = Z \left[\frac{ (\widehat{SR} - SR^{*})\sqrt{n-1}   }{\sqrt{1 - \widehat{\gamma_{3}}\widehat{SR} + \frac{\widehat{\gamma_{4}}-1}{4}\widehat{SR}^{2} } }    \right]  
\end{equation}

Naturally, the estimation of the Sharpe Ratio is subject to significant errors, so the question becomes: "how long should a track record be in order to have statistical confidence that its Sharpe Ratio is above a given threshold?". For a level of certainty $\alpha$, it is shown that the number of observations $n^{*}$ in the track record should be no less than:

\begin{equation}
\label{eq: mTRL}
	n^{*} = 1 + \left[1 - \widehat{\gamma_{3}}\widehat{SR} + \frac{\widehat{\gamma_{4}}-1}{4}\widehat{SR}^{2} \right]\left( \frac{Z_{\alpha}}{\widehat{SR} - SR^{*}}  \right)^{2}.
\end{equation}

\begin{figure}
	\centering
		\includegraphics[width=0.90\columnwidth]{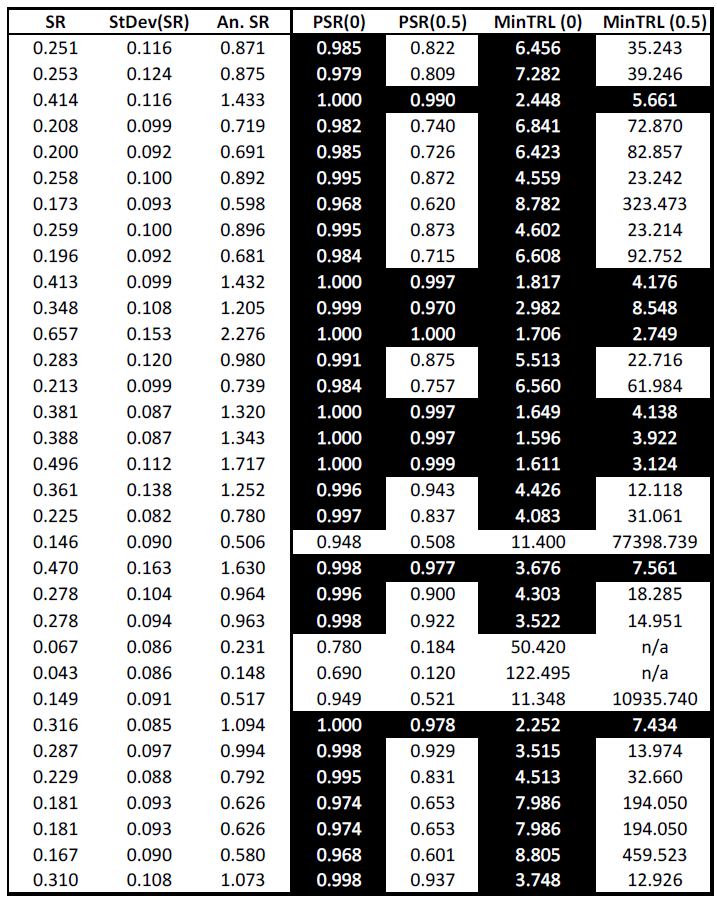}
	\caption{Performance analysis on 33 HFR Indices (Monthly indices). Adapted from~\cite{Bailey:12}}
		\label{fig:fig12Bailey}
\end{figure}

Equation~\ref{eq: mTRL} shows that the exact number of observations will increase as the Sharpe Ratio threshold ($SR^{*}$) gets closer to the observed Sharpe Ratio ($\widehat{SR}$). The more skewed the returns are, or, in other words, the greater the fat tails, the longer the track records must be. An increase in our required level of confidence also results in the need for longer track records. The authors also point that, given the assumptions, the number of required observations ($n^{*}$) should never be less than 30, which means 2.5 years of monthly data, or 0.5769 years of weekly data (52 observations/year), or 0.119 years of daily data (252 observations/year)~\cite{Bailey:12}. In Appendix A, we provide an implementation of Equation~\ref{eq:ProbSR} that calculates the probability of the Sharpe Ratio (PSR) being greater than a given threshold and another implementation of Equation~\ref{eq: mTRL} given the observed time series of returns of the trading algorithm and the desired SR threshold ($SR^{*}$). Both implementation are in MQL 5, the programming language used in the Metatrader platform, which is very similar to C++. In order to illustrate the use of Equations~\ref{eq:ProbSR} and~\ref{eq: mTRL}, we adapted Figure~\ref{fig:fig12Bailey} from the original paper~\cite{Bailey:12} and present the results achieved by the authors. Observe that these equations were applied to 33 HFR indices. HFR indices which are made up of funds managed by human experts, not by autonomous trading strategies. In Figure~\ref{fig:fig12Bailey}, note that only 9 indices presented investment skill over an annualized Sharpe Ratio of 0.5 with a 95\% confidence level (cells in black, PSR(0.5) column), but almost all (29 out of 33) substantiated investment skill over an annualized Sharpe Ratio of zero (PSR(0)). It is also possible to observe the minimum track record ( mTRL) required for each case, considering a Sharpe Ratio of 0.5 ( mTRL(0.5)) or zero ( mTRL(0)). These  mTRL are reported in years, considering monthly observations.

It is important to note that the duration of the track record is dependent on strategy performance itself. So, a fixed duration cannot be stated for any trading strategy. However, it should never shorter than 30 time slices, in our implementation that means at least 30 business days. 

We use the Sharpe Ratio as a measure for strategy performance adjusted to risk, despite the criticism, as we discussed in Section~\ref{sec:sharpe}. It is possible to estimate the probability that the Sharpe Ratio of an autonomous trading strategy is above a given threshold using Equation~\ref{eq:ProbSR}. It is important to note that the higher the threshold the longer the required track record may be. The required track record can be estimated using Equation~\ref{eq: mTRL}. Furthermore, a higher level of certainty also may require a longer track record. Both equations are implemented in the STSE module, which is open source code and publicly available~\cite{stse:21}. After executing the STSE steps, three possible outcomes should be apparent: 

\begin{itemize}
	\item{Probably a bad strategy}: if the track record is long enough and it is \textbf{not} possible to state that the trading strategy's Sharpe Ratio is above  zero or any of the questions 4 to 6 are not answered affirmatively. It is unlikely that is a good trading strategy, at least for the given financial parameters. You should think about changing the parameters or the strategy itself.
	
    \item{Longer track record required}: the track record is \textbf{not} long enough given the observed performance to state with the required level of certainty that strategy performance is above the threshold.  In this situation, the developer should keep evaluating the strategy for more time.
	\item {Perhaps a skillful trading strategy} The track record was long enough; it presented good performance, and questions in Step 2 were answered affirmatively. You should consider testing it in real markets to confirm that performance does not change significantly from simulation. Check Section~\ref{sec:real_markets} for more information.
\end{itemize}

\subsection{Operating in real markets}
\label{sec:real_markets}

We believe that only after thorough evaluation, like that of the STSE method, someone should consider letting real money be managed by a autonomous software agent. Naturally, it is wise to start with a small amount of capital allocated and expand it as the performance persists or improves. The capital investment should be reduced it, if there is a performance degradation. Even with those procedures, it is still true that past returns do not guarantee future ones. As stated by Graham in his famous book~\cite{Graham:06}, investment and speculative operations are different because the first are based on thorough analysis that promises safety of capital and adequate return, while the second lacks theses features. Investors need to be sure which operation they are implementing through their trading robots. The problem of estimating the optimal amount capital that should be allocated to an autonomous trading strategy is a complex question in itself. It may be called \textbf{bet sizing}~\cite{Prado:18}. We intend to address this in the future, but it is out of the scope of this paper.

The STSE method may be used in strategies by yourself or by third parties. However, if dealing with a strategy created by a third party, there may be additional problems that we address in Section~\ref{sec:third_party}.

\subsection{Third-party Strategy Evaluation}
\label{sec:third_party}

The evaluation of a strategy developed by a third-party presents the same challenges a strategy developed personally but with some additional pitfalls. As the trading strategy and its parameters were defined by others, the investor does not know if it suffers from some kind of \textbf{data leakage} (intentional or not) from evaluating data to training data nor does she know how many configurations where tried in order to define the parameter. As pointed out by Bailey and others~\cite{Prado:14}, if one does not know how many trials or parameter configurations were tested, one cannot assess the risk of \textbf{overfitting} and that risk grows with the number of trials. Additionally, any developer is tempted to try the maximum possible number of configurations in order to find the one with the best performance. In fact, even if the developer reveals the number of trials and declares that all required precautions against data leakage were observed, there is still an incentive for him to lie, because the reported performance would be more credible. 

However, even if the number of tested configurations is unknown and the robot's strategy is a 'black box', i.e., there is no access to strategy's source code it is still possible to evaluate the strategy's performance using STSE method. To do so, there must be access to the strategy's trading decisions using reliable live data. This can be done using a trading platform that allows simulated operation with real time data. There are several available like Metatrader~\cite{Metatrader:18}, Interactive Brokers~\cite{IKBR:21},  and Alpaca~\cite{ALPC:21} or others.  It is beyond the scope of this paper to discuss such platforms, but one should be able to use STSE with any platform, since you can extract the equity time series marked to market. The STSE method can also be used no matter the number of assets managed by the strategy, as long as it is possible to obtain the equity time series reliably.

In Section~\ref{sec:results}, we apply the STSE method to several autonomous trading strategies with different financial parameters. We have used the Metatrader platform to perform many experiments. The evaluated strategies include five for which we had access to the source code and five for which we did not have access to source code. They were downloaded from the Metatrader platform and tested. The tested scenarios and results are discussed in the next Section.

\section{STSE Results and Discussion for Several Strategies}
\label{sec:results}

Most of the autonomous trading strategies that we read about and almost all trading strategies available in the Metatrader platform deal with only one asset at time. This is true, despite the fact there are very good reasons for creating strategies that deal with several assets and try to explore their complementarity in terms of risk and return, which is widely used in non-autonomous trading strategies.  We evaluate autonomous trading strategies using one asset per scenario in this Section to reflect the literature. However, that is not a limitation of  STSE method, which can be used for evaluating portfolio strategies.

We present the results achieved by applying the proposed method, Significant Trading-Strategy Evaluation (STSE), to many different trading strategies and assets in this Section. A scenario is defined by an autonomous strategy, a set of assets and a time period. We tested more than 300 scenarios. These many scenarios were splitted into two groups. The first group (FX) focused on Foreign Exchange assets, while the second group (SX) dealt with Stock Exchange assets. We used two different thresholds for the minimum Sharpe Ratio: zero and 0.1, both with a level of confidence of 95\%. These values seem very reasonable but one may set the values in STSE setup that best represents individual preferences.  In FX scenarios, we used five different strategies, four based on technical signals and one based on Machine Learning. We had access to the Source code. However, in SX scenarios we evaluated nine strategies and five of them were developed by third-parties and we did not have access to their source code. We were still able to evaluate them using STSE method as discussed in Section~\ref{sec:third_party}. The two groups of scenarios (FX and SX) and their respective results are described next.

\subsection{Foreign Exchange (FX) Scenarios}
\label{sec:FX}

In the context of Foreign Exchange assets (FX), we executed twenty scenarios composed of five strategies and five different assets, over the period of three years of trading data (Jan-1-2017 to Dec-31-2019). The assets were CFD (Contract for Difference) instruments~\cite{cfd:21}. CFD is a contract that enables two parties to trade based on the  difference between the entry and closing prices. In the case of foreign exchange pairs, a trader may bet that one currency will rise (or fall) in comparison with on other. If she is right, she will profit, and suffer a loss in she is wrong~\cite{cfd:21b}. The foreign exchange pairs are listed in Table~\ref{tab:assets} and one of them (BTCUSD) is based on a cryptocurrency (Bitcoin) against the US dollar. 

We evaluated four well-known strategies: moving average convergence/divergence (\textbf{MACD})~\cite{Katz:00} (pg. 134,150), Moving Average (\textbf{MAMA})~\cite{Katz:00} (pg. 109-132), Moving Average Parabolic Stop and Reverse (\textbf{MAPS})~\cite{Wilder:78} and another version of MAPS with size optimization, that we will call \textbf{MAPS2} from now on. We also used one Trading Strategy based on Random Forest~\cite{castro:21}, that we refer as \textbf{RFOR} from now on. All of them use technical indicators and have some different implementations, each with some parameters. We have used the implementation available in the Metatrader5 platform~\cite{Metatrader:18} with the default parameters defined in Appendix~\ref{sec:setup}. We present each one of the scenarios results next and discuss them in the following subsection. 


\begin{table} [ht]
	\begin{center}
		\begin{tabular}{|| l |l |l  || }
\hline	 \# &	Symbol 	&	 Description		\\
\hline  1 &  EURUSD & Euro vs US dollar \\
\hline  2 & USDJPY & US dollar vs Japan Yen \\
\hline  3 & EURJPY &  Euro vs Japan Yen \\
\hline  4 & GBPUSD	&	 Great Britain Pound vs US dollar \\
\hline  5 & BTCUSD	&	Bitcoin vs US dollar \\
\hline
		\end{tabular}
	\end{center}
	\caption{Selected assets from the most traded assets in Foreign Exchange (Forex) and Cryptocurrency (Bitcoin) Market. }
	\label{tab:assets}
\end{table}

\subsection{Foreign Exchange (FX) Results}

In the next tables, we present the achieved results using the five trading algorithms for each of the five assets. For each simulation defined by a scenario, asset, and trading algorithm, we observed the following values: Return, maximum and minimum quarter returns, Sharpe Ratio, PSR (probability real Sharpe Ratio is over the threshold), and minimum track record length ( mTRL) in number of years for two different levels of Sharpe Ratio thresholds. The first threshold was zero while the second was 0.1 (10\%). The first threshold was significantly easier than the second. We have highlighted the situations where the autonomous traders were able to overcome each respective threshold with black cells and present the achieved results in table~\ref{tab:Results1} for EURUSD pair,table~\ref{tab:Results2} for EURJPY, Table~\ref{tab:Results3} for USDJPY, table~\ref{tab:Results4} for GBPUSD. The Table~\ref{tab:Results5} shows the results for BTCUSD, which is based on the main cryptocurrency, Bitcoin, In this case, there was a high return in the three year period, but also a very negative return in the shorter period. There was a very high volatility for BTCUsD as expected for a cryptocurrency. 




\begin{table} [ht]
	\begin{center}
		\begin{tabular}{|| l |l | l | l | l | l || }
            \hline	     & MACD      & MAMA      & MAPS      & MAPS2 & RFOR \\
\hline   Return &-0.000&0.000&0.000&0.493 & -77.498\\
\hline   Sharpe Ratio&-0.006&0.026&0.026&0.006 & 0.007\\
\hline   PSR(0)&0.180&\BlackCell{0.999}&\BlackCell{0.999}&0.818 & 0.923\\
\hline    mTRL(0)&60.882&\BlackCell{0.798}&\BlackCell{0.816}&67.591 & 0.169\\
\hline   PSR(0.1)&0.000&0.000&0.000&0.000 & 0.000\\
\hline    mTRL(0.1)&0.249&0.107&0.109&0.324 & 0.169\\
            \hline
		\end{tabular}
	\end{center}
	\caption{Results for the five strategies for asset EURUSD in three years. A black cell indicates a situation where the strategy surpasses the proposed Sharpe Ratio threshold: 0 or 0.1. Negative results are possible, because we allowed negative equity (trading with leverage)}
	\label{tab:Results1}
\end{table}

\begin{table} [ht]
	\begin{center}
		\begin{tabular}{|| l |l | l | l | l | l || }
        \hline	     & MACD      & MAMA      & MAPS      & MAPS2  & RFOR \\
\hline   Return (\%) &-0.001&0.000&0.000&-0.886 & -47.451\\
\hline   Sharpe Ratio&-0.011&0.033&0.023&-0.007  & -0.007\\
\hline   PSR(0)&0.070&\BlackCell{1.0}&\BlackCell{0.999}&0.125 & 0.018\\
\hline    mTRL(0)&23.356&\BlackCell{0.411}&\BlackCell{1.033}&42.594 & 11.533\\
\hline   PSR(0.1)&0.000&0.000&0.000&0.000  & 0.000\\
\hline    mTRL(0.1)&0.233&0.109&0.097&0.228  & 0.060\\
        \hline
		\end{tabular}
	\end{center}
	\caption{Results for the five strategies for asset EURJPY in three years. A black cell indicates a situation where the strategy surpasses the proposed Sharpe Ratio threshold:  0 or 0.1}
	\label{tab:Results2}
\end{table}

\begin{table} [ht]
	\begin{center}
		\begin{tabular}{|| l |l | l | l | l | l|| }
		 \hline	     & MACD      & MAMA      & MAPS  & MAPS2 & RFOR \\
\hline   Return (\%) &-0.000&0.001&0.000&-0.856 & -94.526\\
\hline   Sharpe Ratio&-0.006&0.045&0.010&-0.008 & -0.001\\
\hline   PSR(0)&0.194&\BlackCell{1.0}&0.999&0.110 & 0.417\\
\hline    mTRL(0)&68.037&\BlackCell{0.232}&5.006&37.127  & 1173.406\\
\hline   PSR(0.1)&0.000&0.000&0.000&0.000 & 0.000\\
\hline    mTRL(0.1)&0.250&0.163&0.075&0.223 & 0.242\\
        \hline
		\end{tabular}
	\end{center}
	\caption{Results for the five strategies for asset USDJPY in three years. A black cell indicates a situation where the strategs overcome the proposed Sharpe ratio threshold:  0 or 0.1}
	\label{tab:Results3}
\end{table}

\begin{table} [ht]
	\begin{center}
		\begin{tabular}{|| l |l | l | l | l | l || }
        \hline	     & MACD      & MAMA      & MAPS      & MAPS2  & RFOR \\
\hline   Return (\%) &-0.001&0.001&0.003&-0.974 & -88.064\\
\hline   Sharpe Ratio&-0.010&0.033&0.032&-0.013 & -0.008\\
\hline   PSR(0)&0.075&\BlackCell{1.0}&\BlackCell{1.0}&0.023 & 0.043\\
\hline    mTRL(0)&24.667&\BlackCell{0.397}&\BlackCell{0.439}&14.381 & 17.402\\
\hline   PSR(0.1)&0.000&0.000&0.000&0.000 & 0.000\\
\hline    mTRL(0.1)&0.234&0.102&0.106&0.206 & 0.117\\
        \hline
		\end{tabular}
	\end{center}
	\caption{Results for the five strategies for asset GBPUSD in three years. A black cell indicates a situation where the strategy surpasses the proposed Sharpe Ratio threshold:  0 or 0.1}
	\label{tab:Results4}
\end{table}

\begin{table} [ht]
	\begin{center}
		\begin{tabular}{|| l |l | l | l | l | l || }
        \hline	     & MACD      & MAMA      & MAPS      & MAPS2  & RFOR \\
\hline   Return (\%) &0.060&-0.014&-0.014&-99.44& 6478.3\\
\hline   Sharpe Ratio&0.030&-0.030&-0.030&0.040&0.01\\
\hline   PSR(0)&0.943&0.146&0.146&0.932&.753\\
\hline    mTRL(0)&0.490&2.100&2.100&1.100&117.0\\
\hline   PSR(0.1)&0.000&0.000&0.000&0.003&NaN\\
\hline    mTRL(0.1)&0.200&0.100&0.100&0.300&NaN\\
        \hline
		\end{tabular}
	\end{center}
	\caption{Results for the five strategies for asset BTCUSD in three years. A black cell indicates a situation where the strategy surpasses the proposed Sharpe Ratio threshold:  0 or 0.1}
	\label{tab:Results5}
\end{table}

\begin{table} [ht]
	\begin{center}
		\begin{tabular}{|| l |l | l | l | l | l|| }
            \hline	     & MACD      & MAMA      & MAPS      & MAPS2 & RFOR \\
\hline   Return (\%) &-0.000&7.920&6.689&-0.449 & -119.169\\
\hline   Sharpe Ratio&-0.016&0.030&0.017&-0.026  & -0.022\\
\hline   PSR(0)&0.185&0.999&0.977&0.033 & 0.001\\
\hline    mTRL(0)&10.571&0.591&2.122&2.735  & 0.987\\
\hline   PSR(0.1)&0.000&0.000&0.000&0.000 &  0.000\\
\hline    mTRL(0.1)&0.219&0.112&0.105&0.120 &  0.036\\
            \hline
		\end{tabular}
	\end{center}
	\caption{Results for the five strategies for asset EURUSD in six months. A black cell indicates a situation where the strategy surpasses the proposed Sharpe Ratio threshold:  0 or 0.1}
	\label{tab:Results1S}
\end{table}


The results show that none of the tested autonomous traders was able to surpass the 0.5 threshold, but some of them were able to surpass the zero threshold. However, as discussed in Section~\ref{sec:work} and presented in Figure~\ref{fig:fig12Bailey}, some human asset management professionals are able to overcame the 0.5 threshold. In fact, nine out of 33 HFR indices surpassed the 0.5 Sharpe Ratio threshold as presented in~\cite{Bailey:12}. We also tested an easier threshold (0.1) for the selected autonomous, but once again none of them were able to surpass it.

Therefore, we can state that none of the tested strategies showed skill in any of the assets comparable to good human experts. By significant skill, we mean that the strategy performance were above the threshold ( 0 or 0.1) according to the case, with 95\% level of confidence and  mTRL equal or shorter than the simulated period (3 years, in this case). That indicates that creating successful trading strategies may be a really hard task. However, two strategies(MAMA and MAPS) were able to surpass the zero threshold. Even though it is the easiest barrier, it is not completely meaningless. As shown in Figure~\ref{fig:fig12Bailey}, four out of the 33 HFR indices were not able to surpass the zero threshold. RFOR performance was particularly bad. That fact should not be interpreted as a signal that AI-based strategies are not promising, because we used the standard implementation presented in~\cite{castro:21}, which was focused on stock markets rather than FX markets. Furthermore, if we had changed some hyper-parameters or included more information, such as simple moving averages, it is likely its performance would have been enhanced. In fact, we included RFOR to show that the STSE method could also be used with AI-based strategies. 

\begin{table} [ht]
	\begin{center}
		\begin{tabular}{|| l |l | l | l | l | l || }
		 \hline	     & MACD      & MAMA      & MAPS  & MAPS2 & RFOR \\
\hline   Return &-0.000&3.770&0.0&0.055 & -115.144\\
\hline   Sharpe Ratio&-0.014&0.023&N/A&0.005  & -0.018\\
\hline   PSR(0)&0.212&0.999&N/A&0.631  & 0.104\\
\hline    mTRL(0)&13.375&0.716&N/A&82.722  & 5.366\\
\hline   PSR(0.1)&0.000&0.000&N/A&0.000 & 0.000\\
\hline    mTRL(0.1)&0.223&0.071&N/A&0.325  & 0.134\\
        \hline
		\end{tabular}
	\end{center}
	\caption{Results for the five strategies for asset EURJPY in six months. A black cell indicates a situation where the strategy surpasses the proposed Sharpe Ratio threshold:  0 or 0.1}
	\label{tab:Results2S}
\end{table}

We also executed the tests for a much shorter period of time (only six months, from Jul-1, 2019 to Dec-31, 2019). The results are shown below in Tables~\ref{tab:Results1S},~\ref{tab:Results2S},~\ref{tab:Results3S},~\ref{tab:Results4S}, and~\ref{tab:Results5S}. It is possible to observe that MAPS and MAMA strategies were successful only with the USDJPY asset. It indicates that it may be harder for an autonomous strategy to show skill in shorter periods.

\begin{table} [ht]
	\begin{center}
		\begin{tabular}{|| l |l | l | l | l | l || }
        \hline	     & MACD      & MAMA      & MAPS      & MAPS2  & RFOR\\
\hline   Return &-0.000&0.000&0.000&0.553  & -98.774\\
\hline   Sharpe Ratio&-0.014&0.051&0.051&0.025  & 0.003\\
\hline   PSR(0)&0.214&\BlackCell{0.999}&\BlackCell{0.999}&0.920  & 0.579\\
\hline    mTRL(0)&13.581&\BlackCell{0.189}&\BlackCell{0.192}&4.697 & 209.251\\
\hline   PSR(0.1)&0.000&0.000&0.000&0.000 & 0.000\\
\hline    mTRL(0.1)&0.225&0.207&0.214&0.542 & 0.287\\
        \hline
		\end{tabular}
	\end{center}
	\caption{Results for the five strategies for asset USDJPY in six months. A black cell indicates a situation where the strategy surpasses the proposed Sharpe Ratio threshold:  0 or 0.1}
	\label{tab:Results3S}
\end{table}

\begin{table} [ht]
	\begin{center}
		\begin{tabular}{|| l |l | l | l | l | l || }
        \hline	     & MACD      & MAMA      & MAPS      &  MAPS2  & RFOR\\
\hline   Return &-0.000&0.000&4.139&-0.637 & 130.384\\
\hline   Sharpe Ratio&-0.004&0.027&0.017&-0.022 & 0.025\\
\hline   PSR(0)&0.407&0.999&0.977&0.072 & 0.934\\
\hline    mTRL(0)&155.402&0.791&2.124&4.490  & 3.728\\
\hline   PSR(0.1)&0.000&0.000&0.000&0.000  & 0.00\\
\hline    mTRL(0.1)&0.258&0.121&0.105&0.156 & 0.420\\
        \hline
		\end{tabular}
	\end{center}
	\caption{Results for the five strategies for asset GBPUSD in six months. A black cell indicates a situation where the strategy surpasses the proposed Sharpe Ratio threshold:  0 or 0.1}
	\label{tab:Results4S}
\end{table}

\begin{table} [ht]
	\begin{center}
		\begin{tabular}{|| l |l | l | l | l | l || }
        \hline	     & MACD      & MAMA      & MAPS      & MAPS2  & RFOR \\
\hline   Return &0.070&0.010&-0.010&0.000&-192.8\\
\hline   Sharpe Ratio&0.090&0.000&0.000&-1.00&-0.34\\
\hline   PSR(0)&0.947&0.514&0.491&NaN&0.0008\\
\hline    mTRL(0)&0.100&268.800&733.800&NaN&1.5\\
\hline   PSR(0.1)&0.394&0.137&0.125&NaN&0.000\\
\hline    mTRL(0.1)&4.900&0.300&0.300&NaN&0.1\\
        \hline
		\end{tabular}
	\end{center}
	\caption{Results for the five strategies for asset BTCUSD in six months. A black cell indicates a situation where the strategy surpasses the proposed Sharpe Ratio threshold:  0 or 0.1.}
	\label{tab:Results5S}
\end{table}

\subsection{Stock Exchange (SX) Scenarios}
\label{sec:SX}

In addition to crypto and traditional currency scenarios, we have also used STSE to evaluate autonomous trading strategies operating in stock exchange markets. We selected thirty (30) stocks from 3 different stock exchanges, that are listed in Table~\ref{tab:assetsSX}. Nine autonomous trading strategies were used; four of them were also used FX Scenarios (MACD, MAMA, MAPS and MAPS2) and the five others developed by third parties and available for free download in the Metatrader platform in binary distributions without access to their source codes. The complete list is available in Table~\ref{tab:robotsSX}. In the stock exchange scenarios, we executed 270 scenarios using these nine strategies and thirty assets. They were executed using real data from Jan 01, 2018  to Dec 31, 2019.

\begin{table} [ht]
	\begin{center}
		\begin{tabular}{|| l | l || }
\hline	 \# & 	 Strategy  	\\
\hline	1 & MACD 	 	\\
\hline  2 & MAMA 		\\
\hline	3 & MAPSR	 	\\
\hline	4 & MAPS2 	\\
\hline	5 & Dark Venus MT5 (*) \\
\hline	6 & Neural Networks 2 Moving Averages (*)\\
\hline	7 & Dark Moon MT5 (*) \\
\hline	8 & Hyper Trade Global (*) \\
\hline	9 & CAP Random Trader EA MT5 (*) \\

\hline
		\end{tabular}
	\end{center}
	\caption{Selected Trading strategies for Stock Exchange scenarios. The strategies marked (*) were downloaded as binary code from Metatrader platform, without access to source code.}
	\label{tab:robotsSX}
\end{table}

\begin{table}[ht] 
\begin{center}

\begin{tabular}{|| c| c| c | c ||}
\hline \# & Stock Ex.&Asset ID  & Company \\
\hline 1 & Nasdaq&AAPL& Apple Inc. \\
\hline 2 & Nasdaq&AMD& Advanced Micro Devices, Inc. \\
\hline 3 & Nasdaq&MSFT& Microsoft  \\
\hline 4 & Nasdaq& FB& Facebook \\
\hline 5 & Nasdaq&AAL& American Airlines Group, Inc. \\
\hline 6 & Nasdaq&TSL& Tesla, Inc. \\
\hline 7 & Nasdaq&TLRY&  Tilray, Inc. \\
\hline 8 & Nasdaq&V& Visa Inc. \\
\hline 9 & Nasdaq&JNJ& Johnson \& Johnson \\
\hline 10 & Nasdaq&MU&  Micron Technology, Inc. \\
\hline 11 & NYSE& BABA & Alibaba Group  \\
\hline 12 & NYSE&SQ& Square Inc. Cl A \\
\hline 13 & NYSE&BA& Boeing Co. \\
\hline 14 & NYSE&NIO& NIO Inc.  \\
\hline 15 & NYSE&AMC& AMC Entertainment  Inc. \\
\hline 16 & NYSE&JPM& JPMorgan Chase  \\
\hline 17 & NYSE&ABEV&  Ambev  \\
\hline 18 & NYSE&SHOP& Shopify Inc. Cl A \\
\hline 19 & NYSE& AA & Alcoa corp. \\
\hline 20 & NYSE & DI & Didi Inc. \\
\hline 21 & B3&VALE3& Vale \\
\hline 22 & B3&ITUB4& Itaú \\
\hline 23 & B3&PETR4& Petrobras \\
\hline 24 & B3&BBDC4&  Bradesco \\
\hline 25 & B3&B3SA3&  B3\\
\hline 26 & B3&PETR3&  Petrobras\\
\hline 27 & B3&ABEV3&  Ambev\\
\hline 28 & B3& OIBR3&  Oi \\
\hline 29 & B3&ITSA4&  Itausa\\
\hline 30 & B3&BBAS3&  Banco do Brasil\\
\hline 
\end{tabular}
\end{center}
\caption{List of Assets used in SX scenarios}
\label{tab:assetsSX}
\end{table}

\subsection{Stock Exchange (SX) Results}

The SX results are presented in Tables~\ref{tab:data} and~\ref{tab:data2}. The black cells in these tables show the situations where some strategy passed the zero threshold for a given asset. We checked if those strategies with performance above zero, were able to pass the 0.1 threshold and we also checked if some of them could pass an easier limit (0.05). The results are presented in Table~\ref{tab:data3}. Just as was observed in FX scenarios, not a single strategy was able to pass 0.1 threshold or even the easier 0.05 limit. 

\begin{landscape}
\begin{table*}[ht] 
\begin{center}
\resizebox{1.0\textwidth}{!}{
\begin{tabular}{|| c| c| c| c| c| c| c| c| c| c ||}
\hline  &\multicolumn{2}{|c|}{MACD}&\multicolumn{2}{|c|}{MAMA}&\multicolumn{2}{|c|}{MAPS}&\multicolumn{2}{|c|}{MAPS2}\\
\hline Asset ID&PSR(0)& mTRL(0)&PSR(0)& mTRL(0)&PSR(0)& mTRL(0)&PSR(0)& mTRL(0)\\
\hline  AAPL&0.362&8.8&0&0.1&0&0.1&0.004&0.2\\
\hline  AMD&0.218&1.9&0&0.1&0.218&1.9&0.033&0.3\\
\hline  MSFT&0.199&1.6&0&0.1&0&0.1&0&0.1\\
\hline  FB&0.375&11.2&0.012&0.2&0.012&0.2&NaN&NaN\\
\hline  AAL&0.455&88.8&0.434&40.8&0.434&40.8&0.044&0.4\\
\hline  TSLA&0.136&0.5&\BlackCell{0.998}&\BlackCell{0.100}&0.918&0.3&0.001&0.1\\
\hline  TLRY&0.129&0.5&0.918&0.3&0.918&0.3&\BlackCell{0.953}&\BlackCell{0.200}\\
\hline  V&0.308&4.2&0&0.1&0&0.1&0.004&0.2\\
\hline  JNJ&0.668&6.1&0.013&0.2&0.013&0.2&0&0.1\\
\hline  MU&0.422&29.3&0.106&0.7&0.106&0.7&0.001&0.1\\
\hline  BABA&0.348&3.9&0.009&0.1&0.009&0.1&0&0\\
\hline  SQ&0.725&1.7&0.144&0.5&0.144&0.5&0.001&0.1\\
\hline  BA&0.625&11.2&0.075&0.6&0.075&0.6&0.002&0.1\\
\hline  NIO&0.024&0.2&0.473&125.5&0.473&125.5&0.575&16.7\\
\hline  AMC&0.491&1154.9&0.505&3426.3&0.505&3426.3&0.002&0.1\\
\hline  JPM&0.695&4.4&0.003&0.2&0.003&0.2&0&0.1\\
\hline  ABEV&0.297&2.1&0.015&0.1&0.015&0.1&0.713&1.9\\
\hline  SHOP&0.701&2.1&0&0.1&0&0.1&\BlackCell{0.982}&\BlackCell{0.100}\\
\hline  AA&0.8&0.8&0.163&0.6&0.163&0.6&0.417&13.5\\
\hline  DIS&0.608&15.3&0.002&0.1&0.002&0.1&0.009&0.2\\
\hline  VALE3&0.601&20.4&0.304&5.1&0.02&0.3&NaN&NaN\\
\hline  ITUB4&0.567&46.8&0.438&54.9&0.4&21&NaN&NaN\\
\hline  PETR4&0.572&40.7&0.253&3&0.299&4.8&NaN&NaN\\
\hline  BBDC4&0.609&17.6&0.281&4&0.259&3.2&NaN&NaN\\
\hline  B3SA3&0.5&1381322.3&0.262&3.3&0.285&4.2&NaN&NaN\\
\hline  PETR3&0.554&72.2&0.337&7.6&0.259&3.2&NaN&NaN\\
\hline  ABEV3&0.569&43.9&0.546&98.4&0.848&1.3&0.002&0.2\\
\hline  OIBR3&0.585&28.8&0.669&7&0.643&10&NaN&NaN\\
\hline  ITSA4&0.48&546.4&0.37&12.2&0.007&0.2&NaN&NaN\\
\hline  BBAS3&0.601&20.6&0.423&35.2&0.404&22.9&NaN&NaN\\
\hline 
\end{tabular}}
\end{center}
\caption{Four strategies results for thirty SX assets. A black cell indicates a situation where the strategy surpassed the zero Sharpe ratio threshold}
\label{tab:data}
\end{table*}
\end{landscape}


\begin{landscape}
\begin{table*}[ht] 
\begin{center}
\resizebox{1.0\textwidth}{!}{
\begin{tabular}{|| c| c| c| c| c| c| c| c| c| c| c| c ||}
\hline  &\multicolumn{2}{|c|}{Dark Venus MT5}&\multicolumn{2}{|c|}{Neural Networks 2 Moving Averages}&\multicolumn{2}{|c|}{Dark Moon MT5}&\multicolumn{2}{|c|}{Hyper Trade Global}&\multicolumn{2}{|c|}{CAP Random Trader EA MT5}\\
\hline Asset ID&PSR(0)& mTRL(0)&PSR(0)& mTRL(0)&PSR(0)& mTRL(0)&PSR(0)& mTRL(0)&PSR(0)& mTRL(0)\\
\hline  AAPL&0.019&0.3&NaN&NaN&0.585&23.6&0.536&131.2&0.362&8.8\\
\hline  AMD&0.163&1.2&NaN&NaN&0.433&40.6&0.486&867.2&0&0.1\\
\hline  MSFT&0.471&215.8&0.015&0.2&0.427&33.8&0.482&569.9&0&0.1\\
\hline  FB&0.557&55.6&NaN&NaN&0.45&73.1&0.498&79229.1&0.012&0.2\\
\hline  AAL&0.536&142.5&\BlackCell{0.962}&\BlackCell{0.400}&0.478&361.1&0.496&10333.7&0.434&40.8\\
\hline  TSLA&0.503&9710.3&0.471&114.3&0.237&1.2&0.474&143.8&\BlackCell{0.998}&\BlackCell{0.100}\\
\hline  TLRY&0.562&24.8&0.129&0.5&0.51&974.4&0.498&16588.2&0.918&0.3\\
\hline  V&0.522&360.3&NaN&NaN&0.445&57&0.488&1242.2&\BlackCell{1.000}&\BlackCell{0.100}\\
\hline  JNJ&0.204&1.7&NaN&NaN&0.559&52.1&0.488&1258.6&0.013&0.2\\
\hline  MU&0.409&21.8&0.309&4.6&0.466&161.1&0.498&36629.8&0.893&0.7\\
\hline  BABA&0.368&5.2&0.001&0.1&0.421&15&0.483&323.3&\BlackCell{0.991}&\BlackCell{0.100}\\
\hline  SQ&0.532&90.4&0.004&0.1&0.555&30.9&0.486&481.5&0.856&0.5\\
\hline  BA&0.857&1&0.002&0.1&0.534&154.9&0.515&794.3&0.075&0.6\\
\hline  NIO&0.597&9.8&0.834&0.6&0.401&9.5&0.525&147.7&0.473&125.5\\
\hline  AMC&0.549&39.6&\BlackCell{0.990}&\BlackCell{0.100}&0.532&94.1&0.467&83.9&0.491&1288.7\\
\hline  JPM&0.529&221.1&0.042&0.4&0.568&39.1&0.487&1036.2&\BlackCell{0.997}&\BlackCell{0.200}\\
\hline  ABEV&0.435&22.3&\BlackCell{0.999}&\BlackCell{0.100}&0.554&31.4&0.467&85&0.015&0.1\\
\hline  SHOP&0.646&4.3&NaN&NaN&0.605&8.4&0.49&1016.9&0&0.1\\
\hline  AA&0.528&121.7&\BlackCell{0.978}&\BlackCell{0.200}&0.532&93.6&0.47&106.1&0.832&0.6\\
\hline  DIS&0.109&0.8&0.001&0.1&0.556&56.7&0.484&670.6&\BlackCell{0.998}&\BlackCell{0.100}\\
\hline  VALE3&0.009&0.2&NaN&NaN&0.574&38.5&0.736&3.4&0.588&27\\
\hline  ITUB4&\BlackCell{0.975}&\BlackCell{0.400}&NaN&NaN&0.526&318.9&0.679&6.2&0.919&0.7\\
\hline  PETR4&0.646&4.3&NaN&NaN&0.606&18.5&0.607&18.3&0.659&8\\
\hline  BBDC4&0.525&327.8&NaN&NaN&0.5&19249669.2&0.615&15.7&0.583&30.3\\
\hline  B3SA3&NaN&NaN&NaN&NaN&0.501&173919.9&0.536&165.1&0.833&1.4\\
\hline  PETR3&\BlackCell{0.927}&\BlackCell{0.600}&NaN&NaN&0.522&445.3&0.627&12.7&0.5&3048446.5\\
\hline  ABEV3&0.99&0.3&NaN&NaN&0.488&1540.2&0.578&34.2&0.565&50.5\\
\hline  OIBR3&0.422&34.9&NaN&NaN&0.579&34.1&0.513&1190.7&0.246&2.8\\
\hline  ITSA4&0.491&2937.9&NaN&NaN&0.466&189.2&0.567&47.4&\BlackCell{0.967}&\BlackCell{0.4}\\
\hline  BBAS3&0.461&141.4&NaN&NaN&0.552&77.2&0.738&3.3&0.598&21.5\\
\hline 
\end{tabular}}
\end{center}
\caption{Five strategies results for thirty SX assets. A black cell indicates a situation where the strategy surpassed the zero Sharpe ratio threshold}
\label{tab:data2}
\end{table*}
\end{landscape}


\begin{table*}[ht] 
\begin{center}
\resizebox{1.0\textwidth}{!}{
\begin{tabular}{|| c| c| c| c| c| c| c ||}
\hline Asset ID&STRATEGY&PSR(0.05)& mTRL(0.05)&PSR(0.1)& mTRL(0.1)\\
\hline TSLA&  CAP Random Trader EA MT5&0.886&0.4&0.331&3.1\\
\hline V&  CAP Random Trader EA MT5&0.822&1.3&0.053&0.4\\   
\hline BABA&  CAP Random Trader EA MT5&0.800&0.8&0.245&1.2\\
\hline JPM&  CAP Random Trader EA MT5&0.662&6.5&0.030&0.3\\
\hline DIS&  CAP Random Trader EA MT5&0.720&3.4&0.040&0.4\\
\hline ITSA4&  CAP Random Trader EA MT5&0.736&3.4&0.281&4.0\\   
\hline TSLA&  MAMA&0.886&0.4&0.331&3.1\\
\hline TLRY& MAPS2&0.753&1.3&0.377&6.1\\
\hline AAL& Neural Networks 2 Moving Averages&0.439&48.6&0.019&0.3\\
\hline AMC& Neural Networks 2 Moving Averages&0.772&1.1&0.207&0.9\\
\hline ABEV& Neural Networks 2 Moving Averages&0.898&0.4&0.306&2.3\\
\hline AA& Neural Networks 2 Moving Averages&0.702&2.1&0.172&0.7\\
\hline ITUB4& Dark Venus MT5&0.519&586.4&0.013&0.3\\
\hline ABEV3& Dark Venus MT5&0.418&31.6&0.009&0.2\\
\hline SHOP& MAPS2&0.722&1.7&0.176&0.7\\
\hline 
\end{tabular}}
\end{center}
\caption{The strategies that passed the zero threshold results for higher thresholds. A black cell indicates a situation where the strategy surpassed the 0.05 or 0.1 Sharpe ratio threshold}
\label{tab:data3}
\end{table*}

\subsection{Discussion of Results}

In the fifty FX scenarios, we observed that only eleven strategies managed to produce Sharpe Ratio above the zero threshold, and only two strategies (MAMA,MAPS) were able to do so in the six month period. As expected, it is harder for a strategy to prove skill in shorter time periods. None of the strategies were able to pass the 0.1 threshold. The results were similar in the 270 SX scenarios. Only thirteen out of 270 scenarios showed strategies passing the zero Sharpe Ratio threshold. We tested these thirteen strategies against the 0.1 threshold and also against an easier target of 0.05. None of them were able to pass theses thresholds, not even the easier 0.05 limit (see Table~\ref{tab:data3}).

One may argue that such limits are too high. However, it has been observed that professional human portfolio managers are able to pass a much harder threshold, 0.5~\cite{Bailey:12}. In fact, this study, discussed in Section~\ref{sec:work}, showed that twenty-nine (29) out of thirty-three (33) funds were able to pass the zero threshold. Furthermore, nine of them passed the much harder 0.5 threshold. These results seem to indicate that autonomous trading strategies still have a long road ahead to match a human expert's performance. However, it is clear to us that using just one asset for a strategy contributed to the poor performance, because it does not allow risk mitigation through diversification, but as discussed in Section~\ref{sec:pitfalls}, it is common to see autonomous trading strategy focus only on return and disregarding risk mitigation through diversification. The simple use of optimization algorithms, such as quadratic programming~\cite{Brown:12}, and trading with many assets would likely help to improve results. However, it is not clear to us that this would be enough to close the big gap observed among the tested autonomous strategies and experts' performance. 

Naturally, it is possible that autonomous trading strategies dealing with many assets would present much better performance. Nevertheless, the STSE method would still be able to evaluate such multi-asset strategies without any change, as discussed in Section~\ref{sec:third_party}. 

The STSE method was tested on several strategies and assets (foreign currencies, cryptocurrency, and stocks). The strategies were not selected based on high expectation about their performance. They were selected based on the use of several different techniques and availability. We did not engage in a long process to refine the strategy's parameters, which could have helped to improve their performance. Some strategies did surpass the zero threshold, which means that they have an expected Sharpe Ratio higher than zero with 95\% level of confidence and that the observed track record is equal or above the minimum required track record (mTRL). The STSE method open source implementation is freely available in~\cite{stse:21}. In Appendix~\ref{sec:setup}, we detail all simulation setups, and Strategy's parameters used in the tested scenarios. As discussed in Section~\ref{sec:results}, some of the tested strategies are not open source, but all allow binary download free of charge. Using the STSE open source implementation, the same strategies and setup specified in Appendix~\ref{sec:setup}, it is possible to reproduce the results achieved and presented in this paper. In the next and final Section, we provide a brief summary of this study's contribution.  



\section{Conclusions and Future Work}
\label{sec:conc}

In this paper, we presented a method to evaluate autonomous trading strategies regardless of the techniques used for their construction. This method, called STSE, was able to separate skillful trading strategies from those with no skill (expected Sharpe Ratio below zero). We also provided an open source implementation which we believe can be useful for autonomous trading strategy developers. The STSE implementation is freely available at GitHub~\cite{stse:21}. 

We have used STSE to evaluate nine different strategies and thirty five assets including currency, cryptocurrency and stocks. The autonomous strategies include some based on technical analysis (Moving average, MACD and others), machine learning techniques (Random Forest, Neural networks) and some proprietary strategies developed by third-parties that we did not have much information about. We were still able to apply our STSE method to each one of them. These results indicate that the analyzed strategies are far from presenting performance levels comparable to human experts. We also believe that many claims about great performance achieved by some proprietary autonomous strategies may be due to overfitted strategies, since it is hard to avoid data leakage and other common problems. Furthermore, it is important to note that the saying "past returns do not guarantee future returns", is also valid for autonomous trading. Finally, we observe that new approaches, such as deep reinforcement learning and the use of alternative data~\cite{Prado:18}, may be really useful in building better autonomous strategies. However, significant evaluation processes will be needed to tell if you built a really successful strategy or you are just fooling yourself.

\section*{Acknowledgements}
Paulo A.L. Castro is partially funded by CNPq (Brazil) Grant. No. 311838/2017-0.

\bibliographystyle{unsrt}
\bibliography{bibliopa}

\onecolumn

\appendix

\section{Simulations setup}
\label{sec:setup}

There are some parameters that need to be defined in order to execute simulations in the Metatrader platforms. In Table~\ref{tab:parameters}, you will find a list of them, with explanations and the values used in the simulations described in this paper. In Table~\ref{tab:parameters2}, we provide a list of the specific parameters for each Trading agent (or Expert Advisor). The Random Forest trader (RFOR) was used with the basic setup described in~\cite{castro:21}.

\begin{table}[ht]
	\begin{center}
		\begin{tabular}{|| p{4cm} |p{5cm} | p{4cm} || }
\hline	 Name  &	Description &  Value \\
\hline  Expert & Identification of the Trading
Strategy implementation (called Expert Advisor
- EA - in Metatrader) & Several EAs
were used, see Section~\ref{sec:FX} for complete list \\
\hline EA parameters & Specific parameters for each EA & see table~\ref{tab:parameters2} for complete list \\ 
\hline  Symbol & Identification of the target asset & Several assets were used, see Table~\ref{tab:assets} for complete list \\
\hline  Period &  The timeframe for testing/optimization & D1 \\
\hline  Deposit & Initial Balance of EA &  100000 \\  
\hline Currency & Currency used  &  BRL (Brazilian currency, Real) \\
\hline Leverage & Leverage for testing &  1:100 \\
\hline Model & Tick mode & 2 (Open price only ) \\
\hline ExecutionMode & Execution of trade orders with a random delay & 0 (no delay ) \\
\hline Optimization & Genetic optimization of EA parameters & We used both possibilities with and without optimization \\
\hline OptimizationCriterion & Criteria used for optimization & 0 (it means maximize balance value) \\
\hline FromDate & Start date of the simulation & Several values used see in Section~\ref{sec:results} \\
\hline ToDate & End date of the simulation & Several values used see in Section~\ref{sec:results} \\
\hline ForwardMode & Custom mode of forward testing & 1 (Half of the testing period was used for parameter fitting) \\

\hline
		\end{tabular}
	\end{center}
	\caption{Metatrader Trading Agent (Expert Advisor) Parameters  }
	\label{tab:parameters}
\end{table}

\begin{table}[ht] 
	\begin{center}
		\begin{tabular}{|| l |l |  l || }
\hline	 Expert Advisor  &	Parameter name  &  Value \\
\hline  MACD & Inp\_Expert\_Title &  ExpertMACD \\  
\hline  MACD & Inp\_Signal\_MACD\_PeriodFast & 12 \\
\hline  MACD & Inp\_Signal\_MACD\_PeriodSlow & 24 \\
\hline  MACD & Inp\_Signal\_MACD\_PeriodSignal & 9 \\
\hline  MACD & Inp\_Signal\_MACD\_TakeProfit & 50 \\
\hline  MACD & Inp\_Signal\_MACD\_StopLoss & 20 \\
\hline  MAMA & Inp\_Expert\_Title & ExpertMAMA \\
\hline  MAMA & Inp\_Signal\_MA\_Period & 12; 12; 1; 120; N \\
\hline  MAMA & Inp\_Signal\_MA\_Shift & 6; 6; 1; 60; N \\
\hline  MAMA & Inp\_Signal\_MA\_Method & 0; 0; 0; 3; N \\
\hline  MAMA & Inp\_Signal\_MA\_Applied & 1; 1; 0; 7; N \\
\hline  MAMA & Inp\_Trailing\_MA\_Period & 12; 12; 1;120;N \\
\hline  MAMA & Inp\_Trailing\_MA\_Shift & 0;0;1;10;N \\
\hline  MAMA & Inp\_Trailing\_MA\_Method & 0;0;0;3;N \\
\hline  MAMA & Inp\_Trailing\_MA\_Applied & 1;1;0;7;N \\
\hline  MAPS & Inp\_Expert\_Title & ExpertMAPSAR \\
\hline  MAPS & Inp\_Signal\_MA\_Period & 12;12;1;120;Y \\
\hline  MAPS & Inp\_Signal\_MA\_Shift & 6;6;1;60;Y \\
\hline  MAPS & Inp\_Signal\_MA\_Method & 0;0;0;3;Y \\
\hline  MAPS & Inp\_Signal\_MA\_Applied & 0;0;0;6;Y \\
\hline  MAPS & Inp\_Trailing\_ParabolicSAR\_Step & 0.02;0.02;0.002;0.2;Y \\
\hline  MAPS & Inp\_Trailing\_ParabolicSAR\_Maximum & 0.2;0.2;0.02;2.0;Y \\
\hline  MAPS2 & Inp\_Expert\_Title & ExpertMAPSARSizeOptimized \\
\hline  MAPS2 & Inp\_Signal\_MA\_Period & 15;12;1;120;N \\
\hline  MAPS2 & Inp\_Signal\_MA\_Shift & 4;6;1;60;N \\
\hline  MAPS2 & Inp\_Signal\_MA\_Method & 0;0;0;3;N \\
\hline  MAPS2 & Inp\_Signal\_MA\_Applied & 1;1;0;7;N \\
\hline  MAPS2 & Inp\_Trailing\_ParabolicSAR\_Step & 0.02;0.02;0.002000;0.200000;N \\
\hline  MAPS2 & Inp\_Trailing\_ParabolicSAR\_Maximum & 0.2;0.2;0.020000;2.000000;N \\
\hline  MAPS2 & Inp\_Money\_SizeOptimized\_DecreaseFactor & 3;3;0.300000;30.000000;N \\
\hline  MAPS2 & Inp\_Money\_SizeOptimized\_Percent & 10;10;1.000000;100.000000;N \\
\hline
		\end{tabular}
	\end{center}
	\caption{List of Expert Advisors and their specific parameters}
	\label{tab:parameters2}
\end{table}

\end{document}